\documentclass[12pt,preprint]{aastex}

\def\lens{J1632--0033~}
\def\ltsima{$\; \buildrel < \over \sim \;$}
\def\lsim{\lower.5ex\hbox{\ltsima}}
\def\gtsima{$\; \buildrel > \over \sim \;$}
\def\gsim{\lower.5ex\hbox{\gtsima}}
\def\bovera{$S_{\rm B}(\nu)/S_{\rm A}(\nu)$}
\def\covera{$S_{\rm C}(\nu)/S_{\rm A}(\nu)$}

\slugcomment{}

\shorttitle{} 
\shortauthors{Winn, Rusin, \& Kochanek}

\begin{document}

\title{Investigation of the possible third image and\\
mass models of the gravitational lens PMN~J1632--0033}

\author{
Joshua N.\ Winn\altaffilmark{1},
David Rusin,
Christopher S.\ Kochanek
}

\affil{Harvard-Smithsonian Center for Astrophysics, 60 Garden St.,
Cambridge, MA 02138}

\email{jwinn@cfa.harvard.edu; drusin@cfa.harvard.edu;
ckochanek@cfa.harvard.edu}

\altaffiltext{1}{NSF Astronomy \& Astrophysics Postdoctoral Fellow}

\begin{abstract}
We present multi-frequency VLBA\footnote[2]{The Very Long Baseline
Array (VLBA) and Very Large Array (VLA) are operated by the National
Radio Astronomy Observatory, a facility of the National Science
Foundation operated under cooperative agreement by Associated
Universities for Research in Astromomy, Inc.} observations of
PMN~J1632--0033, one of the few gravitationally lensed quasars
suspected of having a central ``odd'' image.  The central component
has a different spectral index than the two bright quasar images.
Therefore, either the central component is not a third image, and is
instead the active nucleus of the lens galaxy, or else it is a third
image whose spectrum is inverted by free-free absorption in the lens
galaxy.  In either case, we have more constraints on mass models than
are usually available for a two-image lens, especially when combined
with the observed orientations of the radio jets of the two bright
quasars.  If there is no third quasar image, the simplest permitted
model is a singular isothermal sphere in an external shear field:
$\beta= 2.05^{+0.23}_{-0.10}$ ($2\sigma$), where $\rho(r) \propto
r^{-\beta}$.  If the central component is a third image, a hypothesis
which can be tested with future high-frequency observations, then the
density distribution is only slightly shallower than isothermal:
$\beta=1.91\pm 0.02$ ($2\sigma$).  We also derive limits on the size
of a constant-density core, and the break radius and exponent of an
inner density cusp.
\end{abstract}

\keywords{gravitational lensing --- galaxies: structure --- galaxies:
nuclei --- quasars: individual (PMN~J1632--0033)}

\section{Introduction}
\label{sec:intro}

The central density profile $\rho(r)$ of galaxies is poorly known, but
is important for understanding galaxy structure and the nature of dark
matter.  Theoretical cold-dark-matter (CDM) models predict a steep
central density cusp, $\rho \propto r^{-\beta}$ with $\beta
\simeq$~1--1.5 \citep[see, e.g.,][]{nfw97,moore99}.  However,
measurements of spiral galaxy rotation curves suggest the dark matter
has a constant-density core.  This has been argued most strongly for
low-surface-brightness galaxies \citep{deblok01} but may hold for all
spiral galaxies \citep{ds00,sb00}, including our own Galaxy
\citep{be01}.  The conflict may be due to observational effects such
as beam smearing, at least in part \citep{vs01}, but the apparent
conflict has encouraged some theorists to invent mechanisms that avoid
producing cusps (e.g., bar-induced evolution, Weinberg \& Katz 2002;
warm dark matter, Bode, Ostriker, \& Turok 2001;
self-interacting dark matter, Spergel \& Steinhardt 2000).

For elliptical galaxies, where rotation curves cannot be measured, the
central density profile is even less well understood.  Observations of
early-type galaxies with the Hubble Space Telescope (HST) show that
the light distribution is cuspy, with cusp parameters that may be
correlated with global galaxy properties \citep{lauer95,faber97}.  But
such studies do not probe mass directly, and are limited to nearby
galaxies due to the requirements on angular resolution.

Multiple-image gravitational lenses can be used to measure the central
surface density of galaxies at significant redshift, by searching for
faint central images of the background object.  In most cases, lenses
should produce an odd number of images \citep{dr80,burke81}, one of
which is faint and appears near the center of the lens galaxy.  This
``central image'' or ``odd image'' corresponds to the central maximum
in the time-delay surface.  However, of the $\sim$80 known systems in
which a galaxy produces multiple images of a quasar or radio source,
almost all consist of either 2 or 4 images.  One system,
APM~08279+5255, definitely has three quasar images \citep{lewis02b},
but the third image may be due to a ``naked cusp'' configuration
rather than a central time-delay maximum \citep{lewis02a}.

No central image has been identified definitively, even among the
radio lenses, where the central image could in principle be seen
through the light and dust of the lens galaxy.  The absence of central
images has been attributed to the demagnification of the image by the
high central surface density of the lens galaxy \citep[see,
e.g.,][]{nsc86,wn93,rm01,eh02,keeton02}.  An ``extra'' radio component
has been detected in several radio lenses, and in three of those cases
it is still unclear whether the extra component is an additional
quasar image or faint radio emission from an active galactic nucleus
(AGN).  Those three systems are 0957+561 \citep{harvanek97},
MG~J1131+0456 \citep{chen93}, and the subject of this paper:
PMN~J1632--0033 \citep{winn02}.

\begin{figure}[h]
\figurenum{1}
\epsscale{0.5}
\plotone{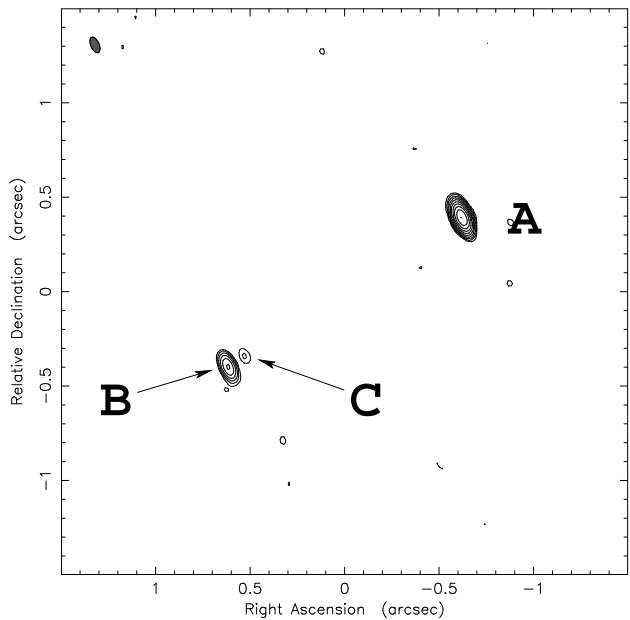}
\caption{ Radio map of \lens reproduced from \citet{winn02},
illustrating the configuration of components A, B, and C.  The map is
based on the 5.0~GHz {\sc merlin} observation of 2001~April~13.  The
restoring beam ($88\times 44$~mas, P.A.\ $24\arcdeg$) is illustrated
in the upper left corner.  Contours begin at $3\sigma$ and increase by
factors of two, where $\sigma=0.13$~mJy~beam$^{-1}$. }
\label{fig:merlin}
\end{figure}

Radio components A and B of J1632--0033 are two images of a $z=3.42$
quasar, but the third, faint component C is of unknown origin (see
Fig~\ref{fig:merlin}).  In an HST image, the lens galaxy appears to be
a fairly circular early-type galaxy with an optical effective radius
of $\approx 0\farcs2$, but it is too faint to be characterized in
detail.  The position of component C is near the optical lens galaxy
position, as expected for either an AGN or a central odd image.  If A,
B, and C are all images of the same quasar, then they should have
similar radio spectra.  However, \citet{winn02} only presented
measurements of the flux density of C at one frequency.

Here we present new VLBA data that, in combination with the previous
data, allow us to measure the flux densities of all three components
at four widely spaced frequencies.  Our analysis shows that C has a
significantly different radio spectrum than A and B.  We argue that
there are two alternative explanations for this difference.  Either C
is not a third quasar image, and instead is the active nucleus of the
lens galaxy; or, C is a third image, but its spectral index is
inverted by free-free absorption in the lens galaxy.

Whichever of these possibilities is true, the additional information
from the VLBA data provides more constraints on mass models of the
lens galaxy than are usually available for a two-image lens.  If the
central component is a third image, then its position and flux are
valuable constraints.  If the central component is the lens galaxy,
then our data provide an unusually precise position for the lens
galaxy, and a strong upper limit on the flux of a third image.  The
VLBA maps also reveal the relative orientations of the radio jets
associated with A and B.  As with all gravitational lensing studies,
we still do not have enough information to determine the complete
density profile of the lens galaxy, but we can use parameterized
models to explore the implications of the data.

Historically, two classes of parameterized models have been used to
study the central-image problem: models with constant-density cores,
and models with central cusps.  Early studies used softened power
laws, $\rho \propto (r^2+r_c^2)^{-\beta/2}$, and found that central
images are suppressed if the core radius $r_c$ is sufficiently small
\citep[see, e.g.,][]{wn93}.  Since the demagnification of the central
image is determined by the central surface density, steeper density
profiles (larger $\beta$) allow for larger core radii for the same
flux limit on the central image.  Power laws shallower than isothermal
($\beta < 2$) produce central images even in the limit $r_c
\rightarrow 0$.

With the gradual realization that theoretical dark-matter profiles and
observed surface-brightness profiles of early-type galaxies have cusps
rather than cores, attention has shifted to the properties of density
distributions with central cusps.  \citet{rm01} used the absence of
odd images in the {\sc class} sample, presently the largest
homogeneous sample of radio lenses, to argue that the global mass
profiles of the lens galaxies cannot be much shallower than
isothermal.  \citet{mkk01} reached a similar conclusion about the
particular radio lens {\sc class}~B1933+503.  \citet{keeton02} found
that the distribution of stars observed in HST images of nearby
early-type galaxies is often sufficiently concentrated to suppress the
central image without recourse to dark matter.

We describe our observations of \lens in \S~\ref{sec:observations},
and our interpretation of the results in \S~\ref{sec:interpretation}.
In \S~\ref{sec:models}, we compute lens models for this system under
the two different hypotheses for the nature of the central component.
We consider density distributions that are scale-free power laws, and
also determine the required properties of a constant-density core or
an inner cusp.  Finally, in \S~\ref{sec:summary}, we summarize our
conclusions and discuss future observations that could test more
definitively whether PMN~J1632--0033 is a three-image quasar.

\section{Observations}
\label{sec:observations}

We observed \lens with the VLBA on 2002~March~14 and 15, for one
eight-hour session each day.  On the first day, the array included one
VLA antenna in place of the VLBA antenna at Pie Town, but on the
second day, the array consisted of the usual ten antennas.  During
both sessions we alternated between observations at the standard
8.4~GHz (3.6~cm) band and the standard 1.7~GHz (18~cm) band.  For both
bands, the observing bandwidth of 32~MHz per polarization was divided
into 4 sub-bands.  Both senses of polarization were recorded with
2-bit sampling.  The data were correlated in Socorro, New Mexico,
producing 16 channels of width 500~kHz from each sub-band, with an
integration time of one second.

Calibration was performed with {\sc aips}\footnote[5]{The Astronomical
Image Processing System ({\sc aips}) is developed and distributed by
the National Radio Astronomy Observatory.} using standard procedures.
We used the observations of \lens to solve for residual delays, rates,
and phases directly (rather than using phase-referencing), with a
fringe-fitting solution interval of 2 minutes.  We used the
multiple-field {\sc clean} deconvolution algorithm available in {\sc
aips} to deconvolve a field centered on A and, simultaneously, a field
of the same size centered on the mid-point between B and C.  For the
8.4~GHz data, the fields were $1024\times 1024$ with a scale of
0.2~mas~pixel$^{-1}$.  For the 1.7~GHz data, the fields were
$512\times 512$ with a scale of 1.0~mas~pixel$^{-1}$.

\begin{figure}
\figurenum{2}
\epsscale{1.0}
\plotone{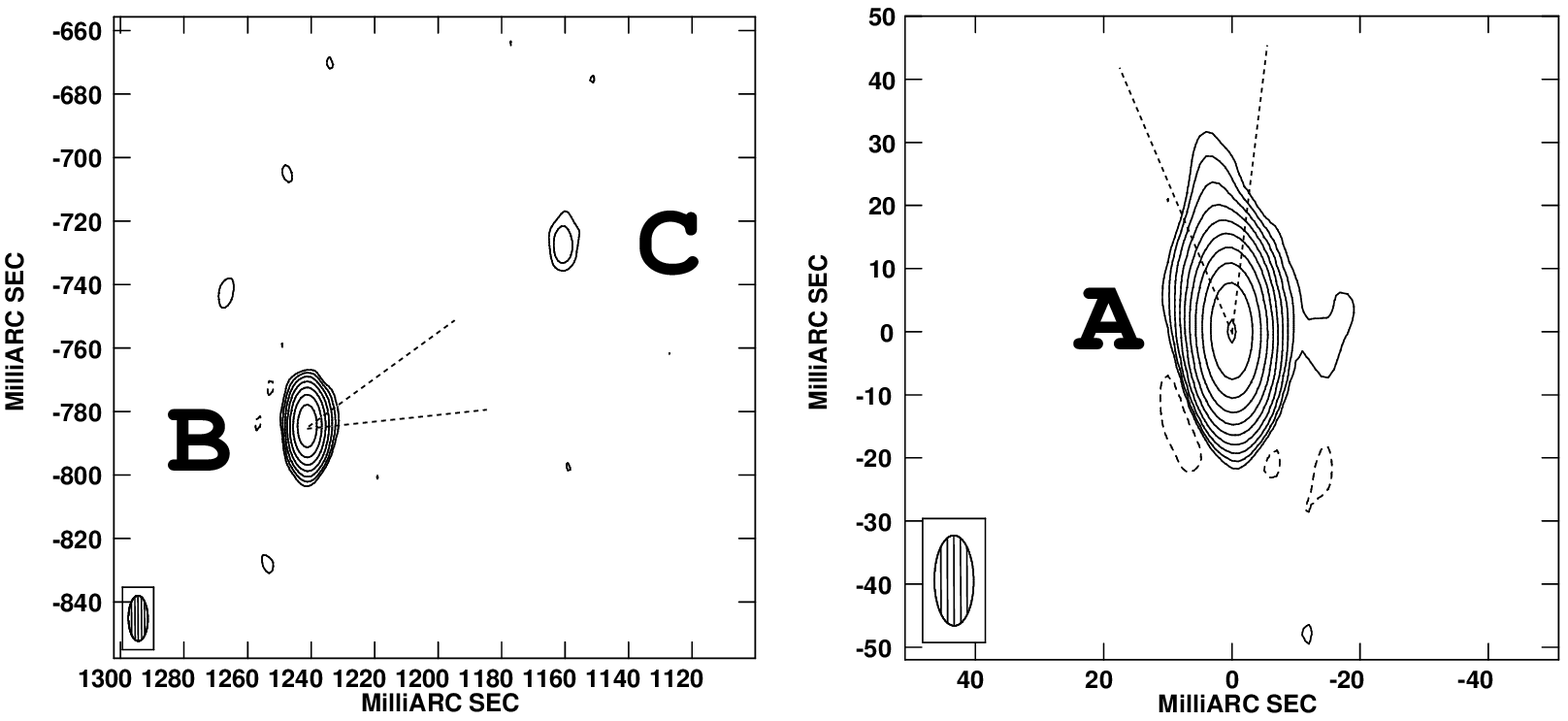}
\caption{ VLBA maps of \lens at 1.7~GHz.  The restoring beam
($14.3\times 6.1$~mas, P.A.~$0\fdg4$) is illustrated in the lower left
corner of each panel.  Contours begin at $3\sigma$ and increase by
factors of two.  The dashed contour is $-3\sigma$.  In the left panel
(components B and C), $\sigma=0.045$~mJy~beam$^{-1}$.  In the right
panel (component A), $\sigma=0.135$~mJy~beam$^{-1}$.  The dotted lines
are the 1$\sigma$ limits we adopted for the jet position angles. }
\label{fig:vlba-18cm}
\end{figure}

\begin{figure}
\figurenum{3}
\plotone{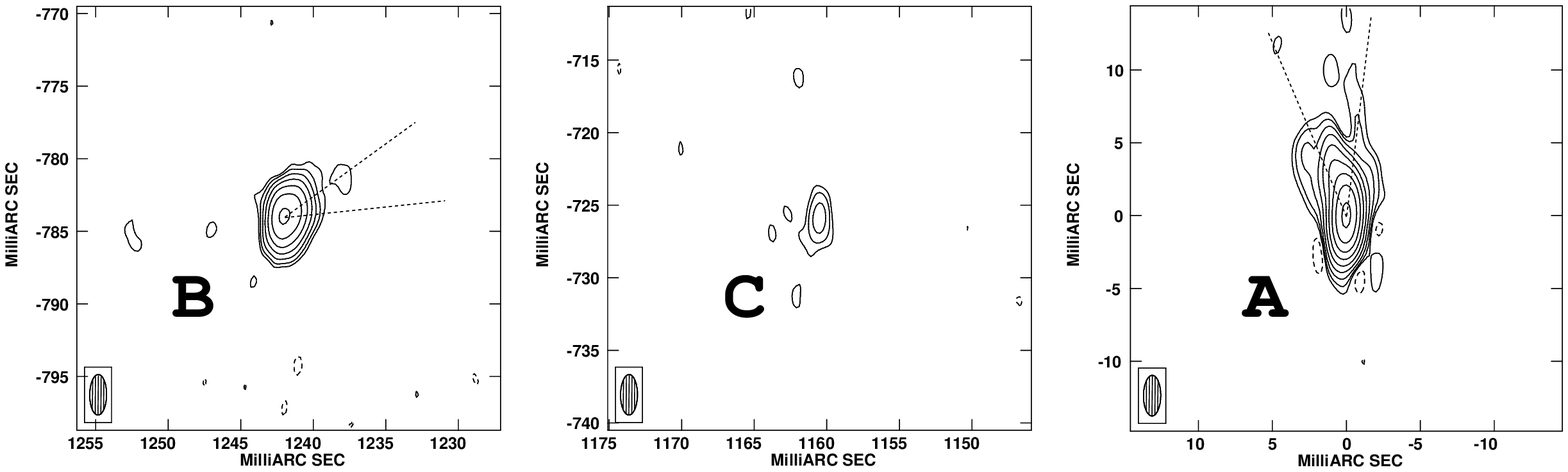}
\caption{ VLBA maps of \lens at 8.4~GHz.  The restoring beam
($2.8\times 1.2$~mas, P.A.~$-0\fdg5$) is illustrated in the lower left
corner of each panel.  Contours begin at $3\sigma$ and increase by
factors of two.  The dashed contour is $-3\sigma$. In the left and
center panels (components B and C), $\sigma=0.040$~mJy~beam$^{-1}$.
In the right panel (component A), $\sigma=0.120$~mJy~beam$^{-1}$.  The
dotted lines are the 1$\sigma$ limits we adopted for the jet position
angles.}
\label{fig:vlba-3.5cm}
\end{figure}

At first, the data from each day were analyzed separately.  Once we
verified that the results of the two sessions were consistent, we
combined the visibility data from each frequency to produce final
maps.  The data were self-calibrated, based on the {\sc clean} model
derived from the preliminary maps, with a 15-second solution interval.
The final 1.7~GHz maps are shown in Fig.~\ref{fig:vlba-18cm}, and the
final 8.4~GHz maps are shown in Fig.~\ref{fig:vlba-3.5cm}.  In the
vicinity of components B and C, the noise level in the maps is close
to the theoretical limit.  However, the noise level is three times
larger in the vicinity of component A due to large sidelobes.  Further
improvement is difficult because of the near-equatorial location of
the object, which results in poor coverage of the visibility plane
even for an eight-hour Earth-rotation synthesis observation.

The flux densities of A, B, and C at each frequency are reported in
Table~\ref{tbl:fluxes}.  Also included in this table are results based
on the previous data of \citet{winn02}, for those cases in which all
three components were detected (or could have been detected, given the
angular resolution).  These previous data were re-analyzed using
natural weighting rather than uniform weighting, to maximize the
sensitivity to the faint component C and to match the present
analysis.  The results were consistent with previous results in all
cases.  Of particular interest was the improved 15~GHz map, in which
component C was detected with flux density $0.84\pm 0.19$~mJy.  This
is consistent with the upper limit of 1.5~mJy given by \citet{winn02},
but the positive detection of C in the lower-noise map provides
another valuable point of comparison.

\begin{deluxetable}{lccccc}
\tabletypesize{\scriptsize}
\tablecaption{Flux density measurements of \lens\label{tbl:fluxes}}
\tablewidth{0pt}
\tablehead{
\colhead{Date} &
\colhead{Observatory} &
\colhead{Frequency} &
\multicolumn{3}{c}{Flux density (mJy)} \\
\colhead{} &
\colhead{} &
\colhead{(GHz)} &
\colhead{A} &
\colhead{B} &
\colhead{C}
}

\startdata
2002~Mar~14-15 & VLBA         & 1.67 & $242\pm 3$ & $16.9\pm 0.1$ & $0.45\pm 0.05$ \\
2000~Apr~29    & VLBA         & 4.99 & $167\pm 2$ & $11.9\pm 0.1$ & $0.60\pm 0.11$ \\
2001~Apr~05    & {\sc merlin} & 4.99 & $207\pm 2$ & $14.2\pm 0.3$ & $0.90\pm 0.11$ \\
2002~Mar~14-15 & VLBA         & 8.42 & $167\pm 2$ & $14.0\pm 0.3$ & $0.73\pm 0.04$ \\
2000~Nov~11    & VLA          & 14.9 & $144\pm 2$ & $ 9.4\pm 0.2$ & $0.84\pm 0.19$ \\
2000~Nov~11    & VLA          & 22.5 & $126\pm 2$ & $ 8.4\pm 0.6$ & $<3.2$ \\
2000~Nov~11    & VLA          & 42.6 & $ 83\pm 1$ & $ 5.4\pm 0.4$ & $<1.8$
\enddata
\end{deluxetable}

For the VLA and {\sc merlin}\footnote[6]{The Multi-Element Radio
Linked Interferometry Network ({\sc merlin}) is a U.K.\ national
facility operated by the University of Manchester at Jodrell Bank
Observatory on behalf of PPARC.} data, the flux densities were
measured by fitting a model consisting of 3 point sources to the
visibility data, using {\sc difmap} \citep{difmap}.  The VLBA data
sets were too large for model-fitting in visibility space.  For these
data, the flux densities were measured in image space with {\sc aips},
using an aperture surrounding each component.  The quoted
uncertainties are the larger of the rms noise and the spread in values
obtained using different choices for the aperture.

With these measurements of the flux densities of all three components
at four widely spaced frequencies, we have the opportunity to test
whether C has the same radio continuum spectrum as A and B.  Assuming
that the flux density of each component obeys $S_i(\nu) \propto
\nu_i^{\alpha}$ (where $\alpha_i$ is the spectral index of component
$i$), a linear least-squares fit to the data in Table~\ref{tbl:fluxes}
gives $\alpha_{\rm A} = -0.27\pm 0.07$, $\alpha_{\rm B} = -0.25\pm
0.07$, and $\alpha_{\rm C} = +0.29\pm 0.18$.  The uncertainties in the
spectral indices include a contribution from the uncertainties in the
absolute flux density scale, which is at least 5\% at all frequencies.

\begin{figure}
\figurenum{4}
\plotone{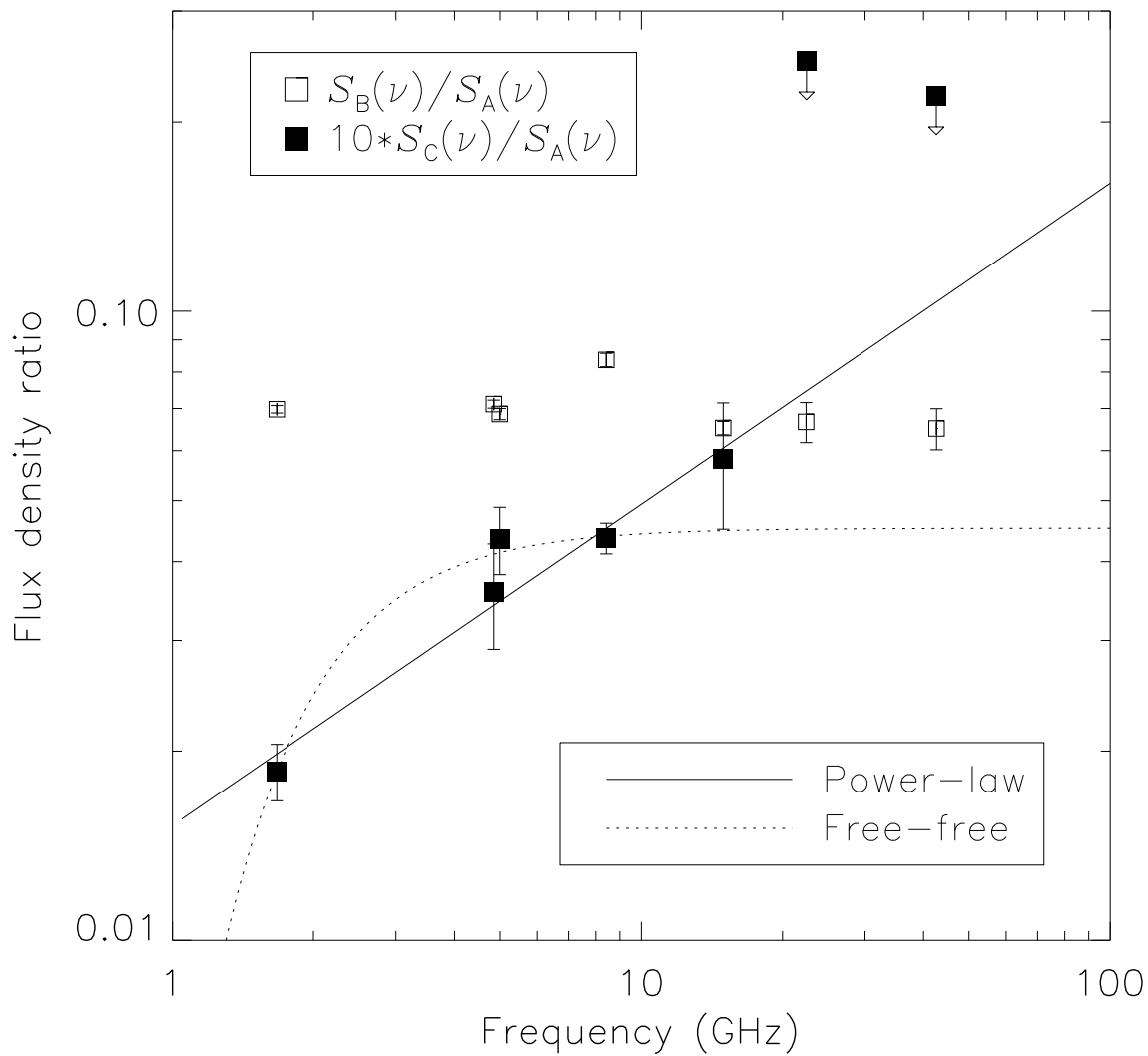}
\caption{ Flux density ratios \bovera~and \covera, as a function of
frequency.  To allow both quantities to be plotted on the same axes,
the values of \covera~have been multiplied by 10.  Upper limits on
\covera~are plotted at 22.5~GHz and 43~GHz.  The straight line is the
best-fitting power law to the \covera~data points at 15~GHz and below.
The dotted line is the best-fitting curve assuming the variation is
entirely due to free-free absorption (see Eq.~\ref{eq:free-free} and
\S~\ref{subsec:propagation}).}
\label{fig:ratios}
\end{figure}

The comparison between $\alpha_{\rm A}$ and $\alpha_{\rm C}$ can be
made more precise by examining the flux density ratios, because the
ratios are independent of the uncertainties in the absolute flux
density scale.  The difference in spectral indices, $\alpha_{\rm C} -
\alpha_{\rm A}$, is the logarithmic slope of \covera.  The flux
density ratios \bovera~and \covera~are plotted in
Fig.~\ref{fig:ratios}.  The solid line represents the best-fit power
law to \covera, which has a slope $\alpha_{\rm C} - \alpha_{\rm A} =
+0.51\pm 0.17$, implying the spectral indices of C and A are
discrepant at the $3\sigma$ level.  The dashed line is a fit to the
data under the hypothesis that C is a third quasar image whose
spectrum had been modified by free-free absorption (see
\S~\ref{sec:interpretation}).  Also plotted are upper limits on
\covera~from VLA data at the two highest frequencies.

The position angles of the jets in A and B were measured by fitting
elliptical components to the VLBA maps with {\sc aips}, using a
procedure (JMFIT) that corrects for the ellipticity of the beam.  For
modeling purposes, we adopted the average of the 1.7~GHz and 8.4~GHz
results, giving $\phi_{\rm A} = 8\arcdeg$ and $\phi_{\rm B} =
-69\arcdeg$, measured east of north.  We estimate the uncertainty in
each position angle to be $\pm 15\arcdeg$, an error range that is
approximately twice the difference between the 1.7~GHz and 8.4~GHz
results.  These ranges of position angle are illustrated by dotted
lines in Figs.~\ref{fig:vlba-18cm} and~\ref{fig:vlba-3.5cm}, although
it is difficult to account for the elliptical beam in a visual
inspection.

\section{The nature of component C}
\label{sec:interpretation}

The flux density ratio \bovera~is nearly independent of observing
frequency, as should be the case for gravitationally lensed images of
a single compact source.  By contrast, the flux density ratio
\covera~increases significantly with frequency.  This immediately
suggests that C is not a third image of the same background source.
Given that its position is consistent with the optical position of the
lens galaxy, it would be natural to conclude that C represents faint
radio emission from the lens galaxy, as discussed further in
\S~\ref{subsec:emission}.  However, there are several mechanisms that
might cause a third quasar image to appear to have a different
spectral index than the two brighter images. These are discussed in
\S~\ref{subsec:variability}--\S~\ref{subsec:propagation}.  The only
mechanism we cannot rule out is free-free absorption
(\S~\ref{subsec:propagation}).

\subsection{Lens galaxy emission}
\label{subsec:emission}

How does the flux density of C compare with the flux density expected
from a typical galaxy at the lens redshift?  Given the radio flux
density and spectral index of component C that were presented in
\S~\ref{sec:observations}, and the lens redshift
estimate\footnote[7]{The lens redshift $z_l=1.0\pm 0.1$ was estimated
by \citet{winn02} by requiring the photometric properties of the lens
galaxy to conform to the fundamental plane of elliptical galaxies, a
method invented by \citet{kochanek00}.} of $z\approx 1$, the implied
1.4~GHz luminosity of the lens galaxy is $1.7\times 10^{24}
h^{-2}$~W~Hz$^{-1}$.

This is significantly more powerful than the radio emission observed
from ordinary galaxies, which is thought to arise from star formation
and is limited to $\lsim 10^{23}$~W~Hz$^{-1}$.  However, it is within
the observed range of luminosities from galaxies harboring active
galactic nuclei (AGN).  For example, \citet{sadler02} identified a
sample of 420 AGN in the 2dF Galaxy Redshift Survey, of which 24\% had
a 1.4~GHz radio luminosity at least as large as the implied luminosity
of component C (after correcting for their different choice of
cosmology).

We conclude that the identification of component C as the active
galactic nucleus of the lens galaxy is a reasonable one.  This would
make \lens unusual among the known lenses, but not unique; the lens
galaxy of B2045+265 has a bright central component that is almost
certainly an active nucleus \citep{fassnacht99}.

\subsection{Variability}
\label{subsec:variability}

If the source is variable, then the instantaneous flux density ratios
between the lensed images will fluctuate, due to the differential time
delays between the images.  If, in addition, the degree of variability
depends on frequency, then the flux density ratios will vary with
frequency.  However, the differential time delay between components B
and C is expected to be very small ($<$0.5~day).  This is much smaller
than typical time scales for quasar variability ($\sim$months).
Indeed, our measurements, which were spaced apart by one day, showed
no sign of variability.  We therefore discount the possibility that
variability is the reason for the discrepancy of spectral indices.

\subsection{Radio substructure}
\label{subsec:substructure}

If the background radio source is extended and has sub-components with
different spectral indices, then these sub-components may each be
magnified by a different factor \citep[see, e.g.,][]{pp99}.  In that
case, the sum of flux densities of the sub-components in a lensed
image would not necessarily have the same spectral index as the
corresponding sum in a different image.

However, in the case of J1632--0033, the source is very compact.  From
Fig.~\ref{fig:vlba-3.5cm} we can limit the separation of
sub-components in image A to $<$5~mas, corresponding to a change in
magnification of only $\Delta\mu\lsim 0.01$ in an isothermal model.
Furthermore, the agreement of $\alpha_{\rm A}$ and $\alpha_{\rm B}$
argues against the presence of a significant effect from
frequency-dependent radio substructure.

\subsection{Resolved-out radio structures}
\label{subsec:resolved}

If a radio source has structure on large angular scales, with Fourier
components corresponding to smaller spatial frequencies than those
measured by an interferometer, then that portion of the radio source
is invisible to the interferometer.  The invisible radio structure is
said to be ``resolved out.''  We must consider the possibility that
the spectral indices of A and C are different only because the
components are being resolved out by different amounts at each
frequency.

We reject this hypothesis based on three facts.  First, there cannot
be much resolved-out structure in components A and B.  During a
monitoring program, we measured their flux densities at 8.4~GHz with
the VLA on 2002~March~14, simultaneous with the VLBA measurement
described in \S~\ref{sec:observations}.  The VLBA flux densities were
$83\pm 7\%$ of the VLA flux densities, despite being measured on
angular scales $\sim$200 times smaller.  Second, if C is a third
quasar image, it is demagnified by a factor of $\sim$10$^2$ and would
have an even smaller fraction of resolved-out structure than A or B.
Third, even if there is resolved-out structure in A, it would be
expected to have a steep radio spectrum, because the extended jets of
quasars generally have steeper spectra than the compact cores.  This
would only worsen the discrepancy between the spectral indices of A
and C, by causing the true spectral index of A to be steeper than
observed.

\subsection{Propagation effects}
\label{subsec:propagation}

Propagation effects that are chromatic, and that affect each lensed
image by a different amount, will cause the spectra of the images to
differ.  For example, reddening by dust causes the optical colors of
lensed images to differ \citep[see, e.g.,][]{motta02,falco99}.  At
radio wavelengths, the propagation effects are due to interstellar
plasma.  The polarizations of lensed images may differ due to Faraday
rotation \citep[see, e.g.,][]{patnaik93}.  Or, in a phenomenon known
as ``scatter-broadening,'' the angular size of a lensed image may be
increased due to scattering by electron-density fluctuations along the
line of sight \citep[see, e.g.][]{jones96}.  However, these two
effects do not alter the spectral indices of lensed images. A possible
exception might arise from scatter-broadening, if it causes a
frequency-dependent amount of flux to be resolved out (see
\S~\ref{subsec:resolved} above), but our data do not show the
characteristic $\nu^{-2}$ dependence of angular size due to
scatter-broadening.

One propagation effect that would alter the spectral indices of lensed
images is free-free absorption by electrons in the lens galaxy.
Free-free absorption decreases strongly with frequency, which would
flatten or invert the spectral index of the central component.  We
must therefore consider the possibility that component C is a third
quasar image whose spectral index has been inverted by free-free
absorption.

Assuming that C is affected by free-free absorption, and A is not, we
would expect the flux density ratio to obey \covera~$=\mu_{\rm
CA}\exp(-\tau_\nu)$, where $\mu_{\rm CA}$ is the lensing magnification
ratio and $\tau_\nu$ is the optical depth to free-free absorption.
Here $\nu=(1+z)\nu_{\rm obs}$ is the frequency at the lens redshift.
A useful approximation for $\tau_\nu$ has been given by
\citet{altenhoff60} and \citet{mh67}:
\begin{equation}
\tau_\nu = 0.08235 \times \left(\frac{T_e}{^{\circ}K}\right)^{-1.35}
                          \left(\frac{\nu}{\rm GHz}\right)^{-2.1}
                          \left(\frac{\int N_e^2 ds}
                             {{\rm pc}\cdot{\rm cm}^{-6}}\right)
         = \left(\frac{\nu}{\nu_c}\right)^{-2.1},
\label{eq:free-free}
\end{equation}
where $T_e$ and $N_e$ are the electron temperature and number density,
and $\nu_c$ is defined implicitly.  This formula predicts a sharp
cut-off at low frequencies, and the asymptotic limit
\covera~$=\mu_{\rm CA}$ as $\nu\rightarrow\infty$.

We solved for the values of the two parameters $\nu_c$ and $\mu_{\rm
CA}$ that provide the best fit to the data.  The results were
$\nu_c=3.2$~GHz ($\nu_{c,{\rm obs}}=1.6$~GHz) and $\mu_{\rm
CA}=0.0045$.  The dotted line in Fig~\ref{fig:ratios} shows the
best-fitting curve, which is statistically consistent with the data
($\chi^2=1.1$ with 3 degrees of freedom).  There are no published
cases of free-free absorption by a lens galaxy for comparison.
However, the best-fitting value $\nu_c=3.2$~GHz is comparable to the
values of $\nu_c=$~1--10~GHz that have been inferred for the central
regions of some active galaxies, using multi-frequency VLBI
measurements of radio jets (for recent examples, see Marr, Taylor, \&
Crawford 2001, Jones et al.\ 2001, Kameno et al.\ 2000, Taylor 1996,
and Levinson, Laor, \& Vermuelen 1995).

Because the data are quantitatively consistent with both the
third-image hypothesis and the free-free absorption hypothesis, we
have only our prior prejudice on the relative likelihood of these two
possibilities to guide us.  One might object that the third-image
hypothesis is contrived, because neither a central image nor free-free
absorption has previously been observed in a radio lens.  However, we
do not consider this ``double novelty'' as a serious objection because
a central image might be expected to suffer more absorption than other
images, due to the potentially large electron column density near the
center of the lens galaxy.  For this reason, we take a neutral
position, and consider the consequences of both hypotheses in the
section that follows.

\section{Updated lens models}
\label{sec:models}

In this section we use information from the VLBA maps to update the
lens models for \lens that were presented by \citet{winn02}.  The goal
of our analysis is to investigate the improvement that results when
two extra pieces of information are available besides the usual image
positions and fluxes: the relative orientations of the radio jets, and
a strong constraint (either a measurement or an upper limit) on the
flux of a central image.  In \S~\ref{subsec:lensgalaxy}, we assume
that component C is due to the lens galaxy.  In \S~\ref{subsec:third},
we explore the consequences if component C is a third quasar image.

\subsection{Case 1: Component C is the lens galaxy}
\label{subsec:lensgalaxy}

In this case, we use the model constraints given in
Table~\ref{tbl:model-constraints-1}.  The position of C is taken to be
the center of the lens galaxy, which is known more precisely than the
lens galaxy position in most other systems.  The upper limit on the
flux density of a third quasar image, expressed as the magnification
ratio $\mu_3/\mu_{\rm A}$, represents the $5\sigma$ limit in the
1.7~GHz VLBA map.  The magnification ratio $\mu_{\rm BA}\equiv
\mu_{\rm B}/\mu_{\rm A}$ is the average of the 15 measurements of
\bovera~presented in this paper and by \citet{winn02}, and the
uncertainty is the variance in those measurements.  In theory, our
error estimates for the positions and magnification ratio should be
enlarged (to $\sim$1-2~mas and 10-20\%, respectively) to account for
possible mass substructure in the lens galaxy \citep[see,
e.g.,][]{mm01,dk02}.  However, in all the cases described below, the
error statistic is dominated by either the third-image flux or the jet
orientations, causing the precise values of the uncertainties on the
positions and magnification ratio to be unimportant.

We define the goodness-of-fit parameter as
\begin{equation}
\label{eq:chisq}
\chi^2 = \sum_{i={\rm A,B}}
    \left[   \frac {({\bf R}'_i - {\bf R}_i)^2}{(\Delta {\bf R}_i)^2}
           + \frac {(\phi'_i-\phi_i)^2}{(\Delta\phi_i)^2} \right]
  + \frac{(\mu'_{\rm BA}-\mu_{\rm BA})^{2}}{(\Delta\mu_{\rm BA})^2},
\end{equation}
in which primed quantities are those determined by the model, and
unprimed quantities are those observed.  In addition, we reject models
that predict a third image brighter than the upper limit.  This is
computationally faster than the more orthodox procedure of adding
another error term to Eq.~\ref{eq:chisq} for the third-image flux, but
we have verified that the results are very nearly equivalent.  This is
because the flux of the third image depends sensitively on the
parameters in our models, causing $\chi^2$ to increase very rapidly
across the boundary in parameter space where the third-image
constraint is violated.

Although in principle there could be a complicated angular structure
in the lens model, due to the ellipticity of the lens galaxy and due
to external perturbations, we do not have enough constraints to
explore such models.  Instead, we use mass models with circular
symmetry, to which we add an external shear field (of strength
$\gamma$ and position angle $\theta_\gamma$) to simulate the combined
effect of galaxy ellipticity and tidal fields from neighboring mass
concentrations.

We begin with global power-law models, for which $\rho(r) \propto
r^{-\beta}$.  The model has 7 free parameters: $\beta$, $\gamma$,
$\theta_\gamma$, the mass normalization, the source location
$(x_s,y_s)$, and the position angle of the source jet $\phi_{\rm s}$.
We applied the 8 constraints of Table~\ref{tbl:model-constraints-1}
using a $\chi^2$-minimization code.  Because the constraint on the
third-image flux is one-sided, and because of the way in which we have
implemented that constraint, the allowed models fit the constraints
perfectly ($\chi^2=0$).  The optimal value is $\beta=2.05$ and the
$2\sigma$ bounds are $1.95 < \beta < 2.28$. This range brackets the
isothermal value $\beta=2$.  The upper bound is enforced by the
observed jet orientations.  The lower bound is enforced by the upper
limit on the flux of the third image.

The lower bound is robust, even if our upper limit on $\mu_3/\mu_{\rm
A}$ is too strong due to the possible extinction of a third image by
free-free absorption, or the lowering of its peak flux density by
scatter-broadening.  For example, if we weaken the limit on the
third-image flux density by a factor of 10 ($\mu_3 / \mu_{\rm A} <
0.01$), the lower bound changes only slightly: $1.91 < \beta < 2.28$.

Interestingly, there is a different (weaker) lower bound on $\beta$
that results from the relative positions of A and B with respect to
the lens center.  For $\beta < 1.83$, the radial critical curve is
sufficiently large to encompass B.  In such models, the parity of B is
reversed, and a brighter third image is predicted in a location
outside the radial critical curve; image B becomes the central odd
image in a three-image system.  Such models can be ruled out.

Next, we investigate the possibility of a constant-density core.
Given the preceding results, we assume that the density profile at
large radii is isothermal and derive an upper bound on the size of a
hypothetical constant-density core.  We adopt a two-dimensional
surface density of the form
\begin{equation}
\label{eq:sis-core}
\kappa(R) = \frac{b}{2\sqrt{R^2 + R_c^2}}, 
\end{equation}
where $R$ is the two-dimensional coordinate.  The constraint on the
third-image flux requires $R_c < 1.6$~mas, corresponding to 0.8\% of
the optical radius of the lens galaxy, or $9h^{-1}$~parsecs at
$z=1$.\footnote{Here and elsewhere, we compute angular-diameter
distances assuming a flat $\Omega_{\rm m}=0.3$ cosmology, giving
$5.61h^{-1}$~kpc per arc second.}

Finally, we explore models with central cusps.  We use the cusped
models introduced by \citet{mkk01}, in which the mass density is
\begin{equation}
\rho(r) =
\frac{\rho_0}{(r/r_b)^\beta [1 + (r/r_b)^2]^{(\eta-\beta)/2} }.
\label{eq:cuspy}
\end{equation}
The density falls off with a logarithmic slope of $\eta$ for $r \gg
r_b$, and $\beta$ for $r \ll r_b$, where $r_b$ is the ``break
radius.''  This model is convenient because the lensing deflection can
be computed analytically.

\begin{figure}
\figurenum{5}
\epsscale{1.0}
\plotone{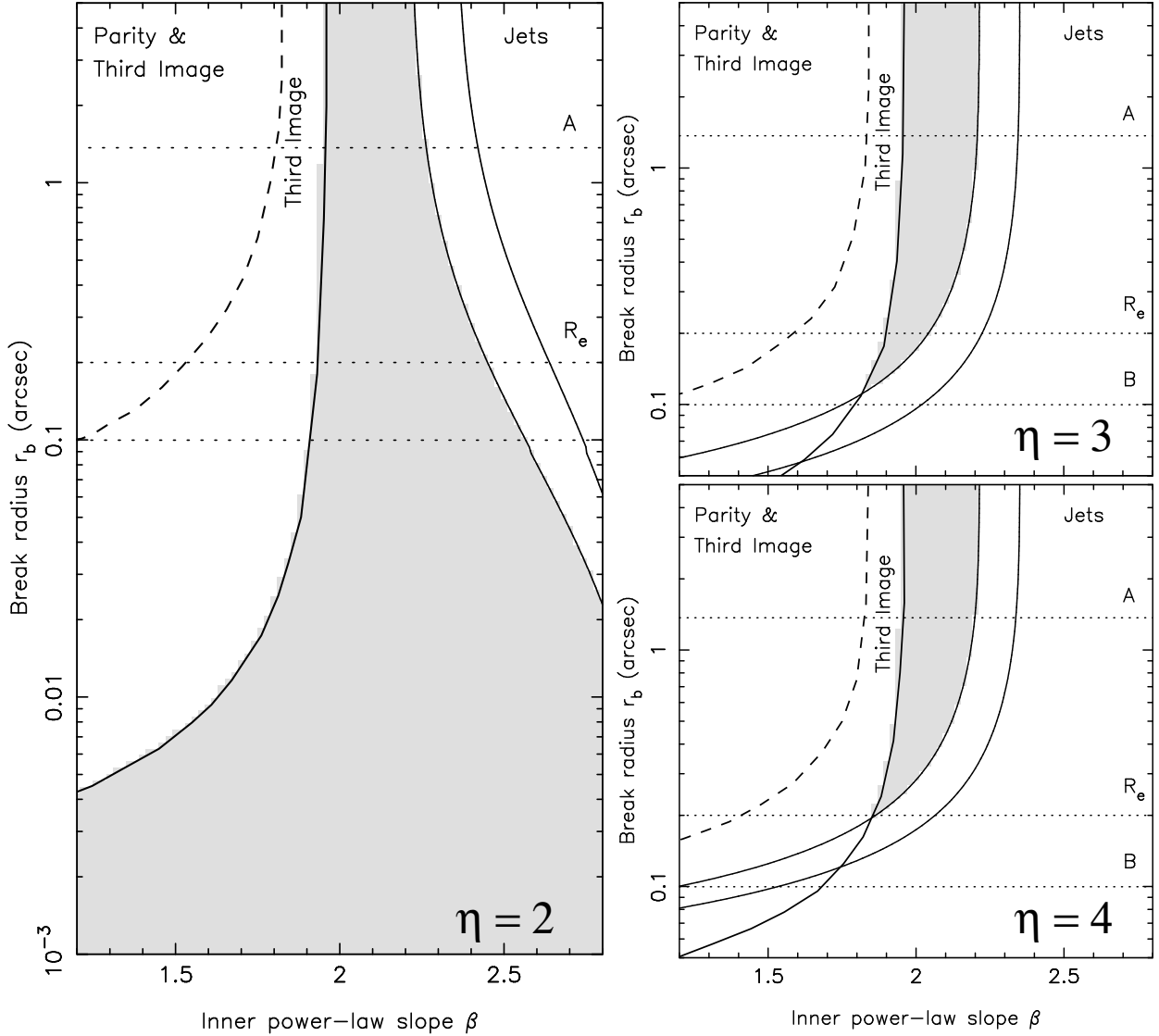}
\caption{Cuspy lens models, assuming that component C is the lens
galaxy.  In all cases, the $1\sigma$ allowed region is shaded.  The
left boundary of the shaded region represents the third-image
constraint. The right boundary of the shaded region is the $1\sigma$
boundary due the jet orientations (the $2\sigma$ boundary is also
shown).  The dashed curve is where image B flips parity.  The
horizontal dotted lines mark the observed radii of images A and B, and
the $I$-band effective radius of the lens galaxy measured by
\citet{winn02}.}
\label{fig:cusp2}
\end{figure}

First, we assume that the mass distribution at large radii is
isothermal ($\eta=2$).  We use the constraints in
Table~\ref{tbl:model-constraints-1} to find the allowed region in the
two-dimensional space spanned by $\beta$ and $r_b$.  As before, the
allowed models fit the constraints perfectly ($\chi^2=0$).  The
results are shown in the left panel of Fig.~\ref{fig:cusp2}.  The
$1\sigma$ allowed region is shaded.  For large $r_b$, the overall mass
distribution must be nearly isothermal, as expected from the preceding
results.  Shallower inner cusps are allowed only when $r_b \lsim
0\farcs02$, corresponding to 10\% of the optical radius, or
$100h^{-1}$~parsecs at $z=1$.  As $r_b$ approaches zero, the
constraint on $\beta$ widens rapidly.  At $r_b\approx 1.6$~mas the
range of allowed exponents encompasses $\beta=0$, which is the limit
of a constant-density core derived in the previous section.

The allowed region of cusp parameters is qualitatively different if we
allow the mass profile of the galaxy at large radii to be steeper than
isothermal.  The upper and lower right panels of Fig.~\ref{fig:cusp2}
show the results for the cases $\eta=3$ and $\eta=4$.  Note that both
NFW \citep{nfw97} and \citet{hernquist90} profiles, with $\beta=1$
central cusps and asymptotic slopes of $\eta=3$ and $\eta=4$
respectively, are not permitted as global models of the mass
distribution, because their inner cusps are too shallow.  Unlike the
case of $\eta=2$, the range permitted for the central cusp exponent
$\beta$ does not widen as $r_b$ shrinks.  The central image and jet
orientation constraints instead pinch off the permitted parameter
space to set a minimum break radius: $r_b>0\farcs12$ ($r_b>0\farcs2$)
for $\eta=3$ ($\eta=4$).  Furthermore, for both $\eta=3$ and $\eta=4$,
the allowed values of $\beta$ bracket the isothermal value $\beta=2$,
regardless of the break radius.  The non-isothermal profile at large
radii must be compensated by a nearly isothermal profile within a
fairly large break radius.  Again, we see that if the mass profile
obeys a power law over the range of radii between images A and B, then
the favored distribution is close to isothermal.

\subsection{Case 2: Component C is a third image}
\label{subsec:third}

For the three-image scenario, we use the model constraints given in
Table~\ref{tbl:model-constraints-2}.  In this case, the lens galaxy
position is taken from the optical measurement of \citet{winn02}, with
error bars enlarged to 23~mas (0.5 pixel) to allow for systematic
errors.\footnote{The smaller error bars of 9~mas quoted by
\citet{winn02} assume a particular form for the point-spread function
and the galaxy profile.  Ordinarily, this would not introduce much
additional error, but in this case component B and the lens galaxy are
both faint and merged in the HST image.}  The value adopted for
\covera~ is based on the fit to the free-free absorption model
described in \S~\ref{subsec:propagation}.  We define the
goodness-of-fit parameter as

\begin{equation}
\chi^2 = \sum_{i={\rm B,C,Gal}} \left[
         \frac {({\bf R}'_i - {\bf R}_i)^2}{(\Delta {\bf R}_i)^2}
         \right]
  + \sum_{i={\rm A,B}} \left[
          \frac{(\phi'_i-\phi_i)^2}{(\Delta\phi_i)^2} \right]
  + \frac{(\mu'_{\rm BA}-\mu_{\rm BA})^2}{(\Delta\mu_{\rm BA})^2}
  + \frac{(\mu'_{\rm CA}-\mu_{\rm CA})^2}{(\Delta\mu_{\rm CA})^2},
\end{equation}
in which primed quantities are those determined by the model, and
unprimed quantities are those observed.

For a scale-free power-law profile in an external shear field, we have
three degrees of freedom.  The model does a good job of reproducing
the configuration, with $\chi^2 = 2.2$.  The logarithmic slope of the
density profile in the optimal model is $\beta = 1.91 \pm 0.02$
($2\sigma$).  If we take the radial density profile to be isothermal
with a constant-density core, as in Eq.~\ref{eq:sis-core}, we can also
achieve a reasonable fit ($\chi^2 = 3.0$), with the optimal value
$r_c= 5.5$~milliarcseconds, or $31h^{-1}$~parsecs, and a $2\sigma$
confidence region of 3.6--7.8~mas.

\begin{figure}
\figurenum{6}
\epsscale{1.0}
\plotone{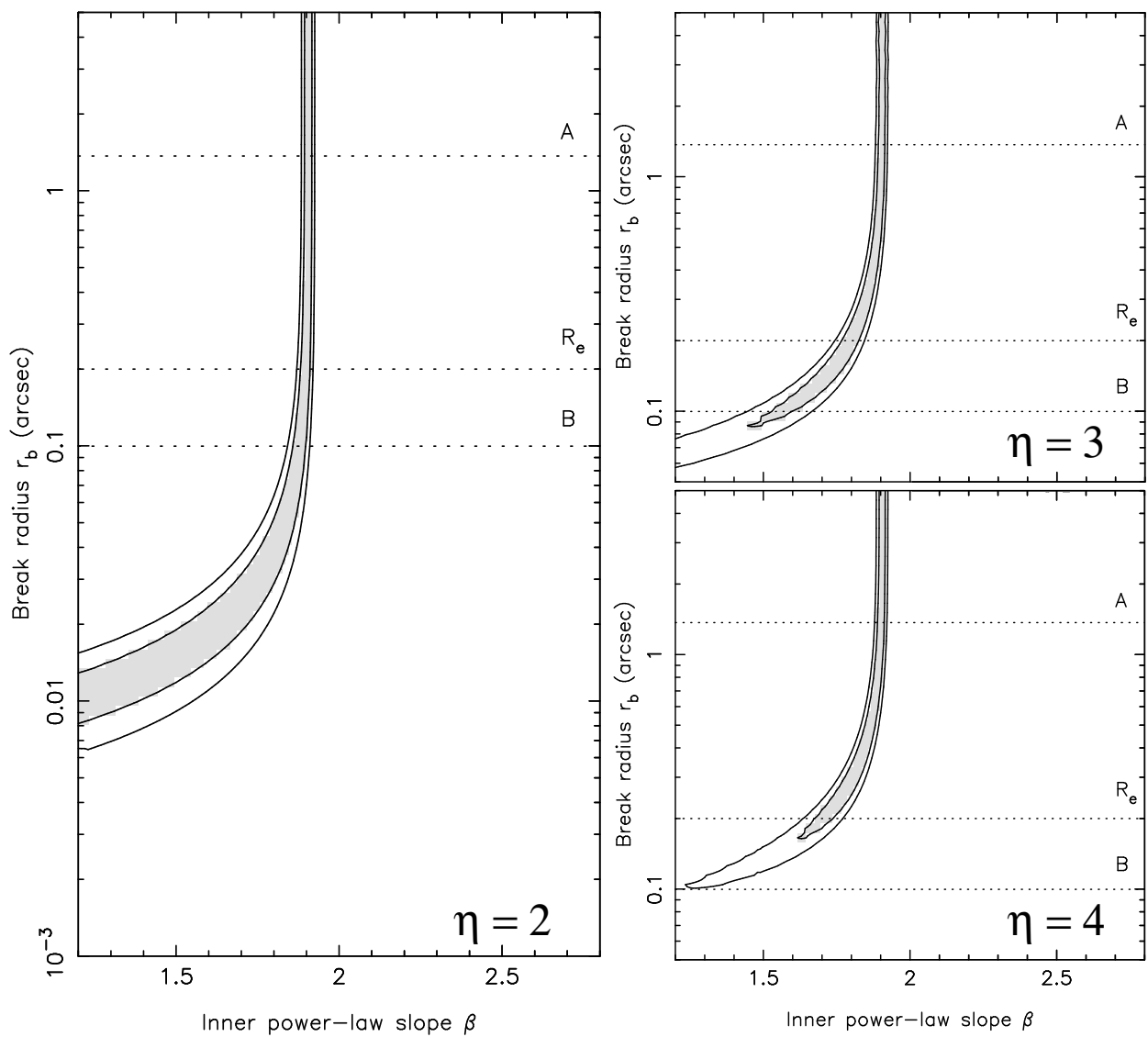}
\caption{Cuspy lens models, assuming that component C is a third
quasar image.  In all cases, the $1\sigma$ allowed region is shaded.
The $1\sigma$ and $2\sigma$ boundaries are marked by solid lines.  The
horizontal dotted lines mark the observed radii of images A and B, and
the $I$-band effective radius of the lens galaxy measured by
\citet{winn02}.}
\label{fig:cusp3}
\end{figure}

When we apply the cuspy lens model of Eq.~\ref{eq:cuspy}, we find that
the allowed region of cusp parameters is much more tightly constrained
than they are in the two-image lens models.  Figure~\ref{fig:cusp3}
shows the results for models in which the radial density distribution
at large radii is isothermal (left panel), or obeys a power law with
logarithmic slope $\eta=3$ (upper right), or $\eta=4$ (lower right).
Both boundaries of the allowed region (shaded) are enforced by the
constraint on the third-image flux.  In particular, for $\beta>2$, no
third image is produced.  For large $r_b$, the cusp must conform to
the value $\beta=1.91$ derived above.  As $r_b$ becomes smaller, the
inner mass distribution must become shallower to compensate for the
steeper mass distribution at large radii.

\section{Discussion and summary of conclusions}
\label{sec:summary}

For the purpose of finding central odd images, the best systems are
radio-loud two-image quasars with a large magnification ratio between
the two bright images.  Radio loudness is important because the radio
components can be observed through the lens galaxy, whereas optical
components are likely to be hidden by the lens galaxy.  Free-free
absorption is potentially significant but can be overcome by observing
at high frequencies.  Two-image systems are better than four-image
systems, as emphasized recently by \citet{keeton02}.  This is because
the projected source position in four-image systems is usually closer
to the lens center, where the central image magnification is smallest.
In addition, four-image systems are subject to a larger magnification
bias than two-image systems \citep{wn93,king96,kks97,rt01,finch02},
leading to a smaller intrinsic source flux and a fainter odd image.
Among two-image systems, the asymmetric doubles are favorable because
the source position is close to the radial caustic, producing the
brightest possible central image for a given mass distribution.

The subject of this paper, PMN~J1632--0033, is exactly this type of
system: an asymmetric two-image radio lens.  For this reason, it
seemed reasonable that the third component found by \citet{winn02}
could be an example of the elusive and long-sought central odd
images. However, we have shown that the third component has a spectral
index that differs by $3\sigma$ from those of the two bright quasar
images.

We have argued that the two most plausible interpretations are that
the central component is emission from the lens galaxy, or else that
it is a third image with the extra complication that its spectral
index has been modified by free-free absorption.  The data are
quantitatively consistent with either scenario.  Lens galaxies are
rarely radio-loud, but the flux density of the component is consistent
with typical radio powers of active galactic nuclei.  No central image
has ever been securely identified for a radio lens, nor has free-free
absorption by a lens galaxy ever been reported (with one possible
exception, {\sc class}~B0128+437; I.\ Browne 2002, private
communication), but one might expect the two phenomena to be related
due to the hot and dense conditions near galaxy centers.

Given the new VLBI data, we have investigated models for the mass
distribution of the lens galaxy under various assumptions.  Whether or
not component C is a third image, we showed that if the mass
distribution obeys a power law $\rho\propto r^{-\beta}$, then it must
be nearly isothermal.  If C is the lens galaxy then
$1.95<\beta<2.28$. If C is a third image then the profile is tightly
constrained to be only slightly shallower: $1.89<\beta<1.93$.  We also
derived limits on a constant-density core radius, or on the break
radius and inner exponent of a central cusp.

These results can be added to the growing collection of evidence that
early-type galaxies have nearly isothermal mass profiles.  In the few
other gravitational lens systems where it is possible to measure the
radial density profile, the results favor isothermal profiles
\citep{kochanek95,cohn01,mkk01, rusin02,kt02}, although there are some
possible counter-examples \citep{tk02,ckh95}.  Moreover, the findings
are consistent with studies of early-type galaxies based on stellar
dynamics \citep[see, e.g.,][]{gerhard01,rix97} and X-ray halos
\citep[see, e.g.,][]{fabbiano89}.

To further elucidate the nature of the central component and the
central density profile of the lens galaxy, it would help to have a
better optical image of the lens galaxy. The current HST image
provides only a weak detection of the lens galaxy.  A better image
could be used to test more precisely whether radio component C is
coincident with the center of the lens galaxy, and to study its
surface brightness profile in detail.  For example, with a better
measurement of the galaxy light profile, one might be able to explore
the balance between the luminous and dark matter in this system.

Even more helpful would be sensitive observations at high radio
frequencies ($\gsim$30~GHz).  As shown in Fig.~\ref{fig:ratios}, the
power-law and the free-free absorption fits diverge significantly at
high frequencies, where free-free absorption becomes negligible.  For
free-free absorption, the ratio \covera~must converge to a constant
value.  For lens galaxy emission, the spectrum need not follow a power
law at the high frequencies, but it would require an unlikely
coincidence for the intrinsic radio spectrum of the lens galaxy to
turn over at high frequencies and become identical to the radio
spectra of A and B.  Observations at low radio frequencies
($\lsim$1~GHz) could also distinguish the hypotheses, in principle,
but it would be harder to achieve the necessary angular resolution and
signal-to-noise ratio.  High frequency observations present their own
challenges, including high noise levels and susceptibility to poor
atmospheric conditions, but the reward would be a definitive discovery
of one of the long-sought but never-seen central images of a
gravitational lens.

\acknowledgments We thank the anonymous referee for constructive
criticism.  We also thank I.\ Browne and L.\ Koopmans for interesting
discussions about radio propagation effects. J.N.W.\ is supported by
an Astronomy \& Astrophysics Postdoctoral Fellowship, under NSF grant
AST-0104347. C.S.K.\ is supported by NASA ATP grant NAG5-9265.

\clearpage

\begin{deluxetable}{lr}
\tabletypesize{\scriptsize}
\tablecaption{Constraints on lens models in which C is the lens
galaxy\label{tbl:model-constraints-1}}
\tablewidth{0pt}

\tablehead{
\colhead{Parameter} & \colhead{Value}
}
\startdata
R.A.(B) -- R.A.(A)        & $+1241.90\pm 0.07$~mas \\
Decl.(B) -- Decl.(A)      & $ -784.00\pm 0.16$~mas \\
R.A.(Gal) -- R.A.(A)      & $+1160.61\pm 0.07$~mas \\
Decl.(Gal) -- Decl.(A)    & $ -726.10\pm 0.16$~mas \\
$\mu_{\rm BA}\equiv\mu_{\rm B}/\mu_{\rm A}$ & $   0.076\pm 0.008$    \\
$\mu_3 / \mu_{\rm A}$     & $ < 0.00093$ \\
$\phi_{\rm A} \equiv$~P.A.\ of A-jet   & $ 8\pm 15\arcdeg$ \\
$\phi_{\rm B} \equiv$~P.A.\ of B-jet   & $-69\pm 15\arcdeg$
\enddata
\end{deluxetable}

\begin{deluxetable}{lr}
\tabletypesize{\scriptsize}
\tablecaption{Constraints on lens models in which C is a third image
\label{tbl:model-constraints-2}}
\tablewidth{0pt}

\tablehead{
\colhead{Parameter} & \colhead{Value}
}
\startdata
R.A.(B) -- R.A.(A)        & $+1241.90\pm 0.07$~mas \\
Decl.(B) -- Decl.(A)      & $ -784.00\pm 0.16$~mas \\
R.A.(C) -- R.A.(A)        & $+1160.61\pm 0.07$~mas \\
Decl.(C) -- Decl.(A)      & $ -726.10\pm 0.16$~mas \\
R.A.(Gal) -- R.A.(A)      & $+1162 \pm 23$~mas \\
Decl.(Gal) -- Decl.(A)    & $ -738 \pm 23$~mas \\
$\mu_{\rm BA}\equiv\mu_{\rm B}/\mu_{\rm A}$ & $   0.076\pm 0.008$    \\
$\mu_{\rm CA}\equiv\mu_{\rm C}/\mu_{\rm A}$ & $   0.0045\pm 0.0009$  \\
$\phi_{\rm A} \equiv$~P.A.\ of A-jet   & $ 8\pm 15\arcdeg$ \\
$\phi_{\rm B} \equiv$~P.A.\ of B-jet   & $-69\pm 15\arcdeg$
\enddata
\end{deluxetable}

\end{document}